\documentclass[12pt]{article}
\usepackage{graphicx}
\begin{document}

\title{\huge Rotations in the Space of Split Octonions }

\author{{\bf Merab Gogberashvili} \\ \\
{\it Andronikashvili Institute of Physics}\\
6 Tamarashvili Street, Tbilisi 0177, Georgia\\and \\
{\it Javakhishvili State University} \\ 3 Chavchavadze Avenue,
Tbilisi 0128, Georgia \\ \\ {E-mail: gogber@gmail.com }}
\date{Received: \today}
\maketitle
\begin{abstract}
The geometrical application of split octonions is considered. The
modified Fano graphic, which represents products of the basis
units of split octonionic, having David's Star shape, is
presented. It is shown that active and passive transformations of
coordinates in octonionic '8-space' are not equivalent. The group
of passive transformations that leave invariant the norm of split
octonions is $SO(4,4)$, while active rotations is done by the
direct product of $O(3,4)$-boosts and real non-compact form of the
exceptional group $G_2$. In classical limit these transformations
reduce to the standard Lorentz group.
\end{abstract}
\medskip


\section{Introduction}

Non-associative algebras may surely be called beautiful
mathematical entities. However, they have never been
systematically utilized in physics, only some attempts have been
made toward this goal. Nevertheless, there are some intriguing
hints that non-associative algebras may play essential role in the
ultimate theory, yet to be discovered.

Octonions are one example of a non-associative algebra. It is
known that they form the largest normed algebra after the algebras
of real numbers, complex numbers, and quaternions \cite{Sc}. Since
their discovery in 1844-1845 by Graves and Cayley there have been
various attempts to find appropriate uses for octonions in physics
(see reviews \cite{Oct}).  One can point to the possible impact of
octonions on: Color symmetry \cite{Color}; GUTs \cite{GUT};
Representation of Clifford algebras \cite{Cliff}; Quantum
mechanics \cite{QM}; Space-time symmetries  \cite{Rel}; Field
theory \cite{QFT}; Formulations of wave equations \cite{WE};
Quantum Hall effect \cite{Hall}; Kaluza-Klein program without
extra dimensions \cite{KK}; Strings and $M$-theory \cite{String};
{\it etc}.

In this paper we would like to study rotations in the model where
geometry is described by the split octonions \cite{Go}.


\section{Octonionic Geometry}

Let us review the main ideas behind the geometrical application of
split octonions presented in our previous papers \cite{Go}. In
this model some characteristics of physical world (such as
dimension, causality, maximal velocities, quantum behavior, etc.)
can be naturally described by the properties of split octonions.
Interesting feature of the geometrical interpretation of the split
octonions is that, in addition to some other terms, their norms
already contain the ordinary Minkowski metric. This property is
equivalent to the existence of local Lorentz invariance in
classical physics.

To any real physical signal we correspond 8-dimensional number,
the element of split octonions,
\begin{equation} \label{s}
s = ct  + x^nJ_n + \hbar \lambda^nj_n + c\hbar\omega I~, ~~~~~(n =
1, 2, 3)
\end{equation}
where we have one scalar basis unit (which we denote as $1$), the
three vector-like objects $J_n$, the three pseudovector-like
elements $j_n$ and one pseudoscalar-like unit $I$. The eight real
parameters that multiply basis elements in (\ref{s}) we treat as
the time $t$, the special coordinates $x^n$, some quantities
$\lambda^n$ with the dimensions momentum$^{-1}$ and the quantity
$\omega$ having the dimension energy$^{-1}$. We suppose also that
(\ref{s}) contains two fundamental constants of physics - the
velocity of light $c$ and the Planck constant $\hbar$.

The squares of basis units of split octonions are inner product
resulting unit element, but with the opposite signs,
\begin{equation} \label{JjI}
J_n^2=1~,~~~~~ j_n^2=-1~, ~~~~~ I^2=1~.
\end{equation}
Multiplications of different hyper-complex basis units are defined
as skew products and the algebra of basis elements of split
octonions can be written in the form:
\begin{eqnarray} \label{algebra}
J_nJ_m &=& - J_mJ_n = \epsilon_{nmk} j^k~, \nonumber\\
j_nj_m &=& -j_mj_n = \epsilon_{nmk} j^k~, \nonumber\\
J_nj_m &=& - j_mJ_n = - \epsilon_{nmk}J^k~, \\
J_nI &=& - IJ_n = j_n~, \nonumber \\
j_nI &=& -Ij_n = J_n~, \nonumber
\end{eqnarray}
where $\epsilon_{nmk}$ is the fully antisymmetric tensor $(n,m,k =
1,2,3)$.

From (\ref{algebra}) we notice that to generate complete basis of
split octonions the multiplication and distribution laws of only
three vector-like elements $J_n$, which describe special
directions, are needed. In geometrical application this can
explain why classical space has three dimensions. The three
pseudovector-like basis units $j_n$ can be defined as the binary
products
\begin{equation} \label{jI}
j_n = \frac{1}{2} \epsilon_{nmk}J^mJ^k~,
\end{equation}
and thus can describe oriented orthogonal planes spanned by two
vector-like elements $J_n$. The seventh basic unit $I$ (oriented
volume) is formed by the products of all three fundamental basis
elements $J_n$ and has three equivalent representation,
\begin{equation} \label{I}
I = J_1j_1 =  J_2j_2 = J_3j_3 ~.
\end{equation}

Multiplication table of octonionic units is most transparent in
graphical form. To visualize the products of ordinary octonions
usually the Fano triangle is used, where the seventh basic unit
$I$ is place at the center of the graphic. In the algebra of split
octonions we have less symmetry and for proper description of the
products (\ref{algebra}) the Fano graphic should be modified by
shifting $I$ from the center of the Fano triangle. Also we shall
use three equivalent representations of $I$, (\ref{I}), and,
instead of Fano triangle, we arrive to King David's shape duality
plane for products of split octonionic basis elements:
\begin{figure}[ht]
\begin{center}
\includegraphics[width=10cm]{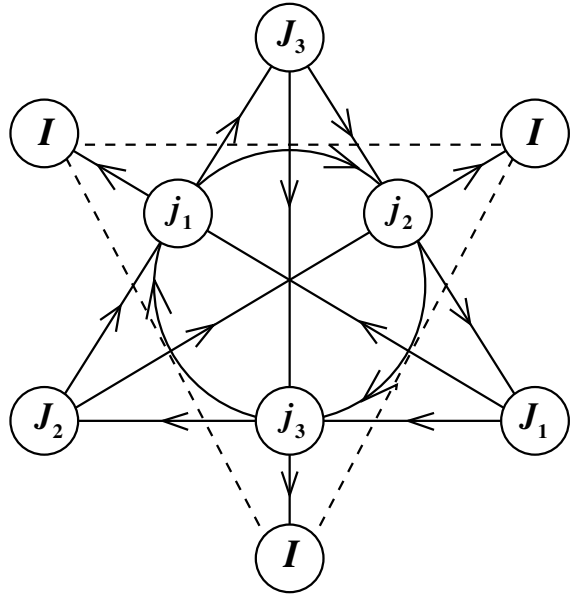}
\caption{\it Display of split octonion multiplication rules
David's Star Plane}
\end{center}
\end{figure}

\noindent On this graphic the product of two basis units is
determined by following the oriented solid line connecting the
corresponding nodes. Moving opposite to the orientation of the
line contribute a minus sign in the result. Dash lines on this
picture just show that the corners of the triangle with $I$ nodes
are identified. Note that products of triple of basis units lying
on a single line is associative and not lying on a single line are
precisely anti-associative.

Conjugation, what can be understand as a reflections of the
vector-like basis units $J_n$, reverses the order of octonionic
basis elements in any given expression and thus
\begin{eqnarray} \label{bar}
J_n^+ = - J_n ~, \nonumber \\
j_n^+ = \frac 12\epsilon _{nmk}(J^mJ^k)^+ = \frac 12 \epsilon
_{nmk}J^{k+}J^{m+} = - j_n~, \\
I^+ = (J_nj_n)^+ = j_n^+J_n^+= - I~,\nonumber
\end{eqnarray}
there is no summing in the last formula. So the conjugation of
(\ref{s}) gives
\begin{equation} \label{s*}
s^+ = ct - x_nJ^n - \hbar\lambda_nj^n - c\hbar\omega I~.
\end{equation}
Using the expressions (\ref{JjI}) one can find that the norm of
(\ref{s}),
\begin{equation} \label{sN}
s^2 = ss^+ = s^+s = c^2t^2 - x_nx^n + \hbar^2 \lambda_n\lambda^n -
c^2\hbar^2\omega^2 ~,
\end{equation}
has $(4+4)$ signature. If we consider $s$ as the interval between
two octonionic signals we see that (\ref{sN}) reduce to the
classical formula of Minkowski space-time in the limit $\hbar
\rightarrow 0$.

Using the algebra of basis elements (\ref{algebra}) the octonion
(\ref{s}) can be written in the equivalent form:
\begin{equation} \label{s'}
s = c(t + \hbar \omega I) +  J^n(x_n + \hbar \lambda_n I) ~.
\end{equation}
We notice that the pseudoscalar-like element $I$ introduces the
'quantum' term corresponding to some kind of uncertainty of
space-time coordinates. For the differential form of (\ref{s'})
the invariance of the norm gives the relation:
\begin{equation} \label{ds/dt}
\frac{ds}{dt} = \sqrt{ \left[ 1 - \hbar^2
\left(\frac{d\omega}{dt} \right)^2\right] - \frac{v^2}{c^2} \left[
1 - \hbar^2 \frac{d\lambda^n}{dx^m}\frac{d\lambda_n}{dx_m}\right] } ~,
\end{equation}
where $v_n = dx_n /dt$ denotes 3-dimensional velocity measured in
the frame (\ref{s}). So the generalized Lorentz factor
(\ref{ds/dt}) contains an extra terms and the dispersion relation
in our model has a form similar to that of double-special
relativity models \cite{double}. Extra terms in (\ref{ds/dt}) go
to zero in the limit $\hbar \rightarrow 0$ and ordinary Lorentz
symmetry is restored.

From the requirement to have the positive norm (\ref{sN}) from
(\ref{ds/dt}) we obtain several relations
\begin{equation} \label{delta}
v^2 \leq c ^2 ~, ~~~~~\frac{dx^n}{d \lambda^n} \geq \hbar
~,~~~~~\frac{dt}{d\omega} \geq \hbar ~.
\end{equation}
Recalling that $\lambda$ and $\omega$ have dimensions of
momentum$^{-1}$ and energy$^{-1}$ respectively, we conclude that
Heisenberg uncertainty principle in our model has the same
geometrical meaning as the existence of the maximal velocity in
Minkowski space-time.


\section{Generalized Lorentz transformations}

To describe rotations in 8-dimensional octonionic space (\ref{s})
with the interval (\ref{sN}) we need to define exponential maps
for the basis units of split octonions.

Since the squares of the pseudovector-like elements $j_n$, as it
is for ordinary complex unit, is negative, $j_n^2 = -1$, we can
define
\begin{equation} \label{R(j)}
e^{j_n\theta_n } = \cos \theta_n + j_n\sin \theta_n ~,
\end{equation}
where $\theta_n$ are some real angles.

At the same time for the other four basis elements $J_n, I$, which
have the positive squares $J_n^2 = I^2 = 1$, we have
\begin{eqnarray} \label{R(J,I)}
e^{J_n m_n } &=& \cosh m_n + j_n\sinh m_n ~, \nonumber \\
e^{I \sigma } &=& \cosh \sigma + I \sinh \sigma ~,
\end{eqnarray}
where $m_n$ and $\sigma$ are real numbers.

In 8-dimensional octonionic 'space-time' (\ref{s}) there is no
unique plane orthogonal to a given axis. Therefore for the
operators (\ref{R(j)}) and (\ref{R(J,I)}) it is not sufficient to
specify a single rotation axis and an angle of rotation. It can be
shown that the left multiplication of the octonion $s$ by one of
the operators (\ref{R(j)}), (\ref{R(J,I)}) (e.g. $\exp
(j_1\theta_1)$) yields four simultaneous rotations in four
mutually orthogonal planes. For the simplicity we consider only
the left products since it is known that for octonions one side
multiplications generate all the symmetry group that leave
octonionic norms invariant \cite{Man-Sch}.

So rotations naturally provide splitting of a octonion in four
orthogonal planes. To define these planes note that one of them is
formed just by the hyper-complex element that we choice to define
the rotation ($j_1$ in our example), together with the scalar unit
element of the octonion. Other three orthogonal planes are given
by the three pairs of other basis elements that lay with the
considered basis unit ($j_1$ in the example) on the lines emerged
it in the David's Star (see Figure 1). Thus the pairs of basis
units that are rotate into each other are just the pairs products
of which give considered basis unit and thus form an associative
triplets with it. For example, the basis unit $j_1$, according to
Figure 1, have three different representations in the octonionic
algebra,
\begin{equation} \label{j1}
j_1 = J_2J_3 = j_2j_3 = J_1I~.
\end{equation}
Than orthogonal to $(1 - j_1)$  planes are $(J_2 - J_3), (j_2 -
j_3)$ and $(J_1 - I)$. Using (\ref{j1}) and the representation
(\ref{R(j)}) its possible to 'rotated out' four octonionic axis
and (\ref{s}) can be written in the equivalent form
\begin{equation} \label{s-j1}
s = N_t e^{j_1 \theta_t} +  N_x e^{j_1 \theta_x}J_3
+ N_\lambda e^{j_1 \theta_\lambda}j_2 + N_\omega e^{j_1 \theta_\omega}I ~,
\end{equation}
where
\begin{eqnarray} \label{N-j1}
N_t &=& \sqrt{c^2t^2 + \hbar^2\lambda_1^2} ~, ~~~~~~~
N_x = \sqrt{x_2^2 + x_3^2} ~, \nonumber \\
N_\lambda &=& \hbar \sqrt{\lambda_2^2 + \lambda_3^2} ~, ~~~~~
N_\omega = \sqrt{x_1^2 + c^2\hbar^2\omega^2} ~,
\end{eqnarray}
are the norms in four orthogonal octonionic planes and the angles
are done by:
\begin{eqnarray}
\theta_t &=& \arccos (t/N_t) ~, ~~~~~~~~~
\theta_x = \arccos (x_3 / N_x) ~, \nonumber \\
\theta_\lambda &=& \arccos (\hbar \lambda_2/N_\lambda) ~, ~~~~~
\theta_\omega = \arccos (c\hbar \omega / N_\omega) ~. \nonumber
\end{eqnarray}
This decomposition of split octonion is valid only if the full
norm of the octonion (\ref{sN}) is positive, i.e.
\begin{equation} \label{s>0}
s^2 = N_t^2 - N_x^2 + N_\lambda^2 - N_\omega^2 > 0~.
\end{equation}
Similar to (\ref{s-j1}) decomposition exists for the other two
pseudovector-like basis units $j_2$ and $j_3$.

In contrast with uniform rotations giving by the operators $j_n$
we have limited rotations in the planes orthogonal to $(1-J_n)$
and $(1-I)$. However, we can still perform similar to (\ref{s-j1})
decomposition of $s$ using expressions of the exponential maps
(\ref{R(J,I)}). But now, unlike to (\ref{N-j1}), the norms of
corresponding planes are not positively defined and, instead of
the condition (\ref{s>0}), we should require positiveness of the
norms of each four planes. For example, the pseudoscalar-like
basis unit $I$ have three different representations (\ref{I}) and
it can provide the hyperbolic rotations (\ref{R(J,I)}) in the
orthogonal planes $(1-I), (J_1 - j_1), (J_2 - j_2)$ and $(J_3 -
j_3)$. The expressions for the 2-norms (\ref{N-j1}) in this case
are: $c \sqrt{t^2 - \hbar^2 \omega^2}, \sqrt{x_1^2 - \hbar^2
\lambda_1^2}, \sqrt{x_2^2 - \hbar^2 \lambda_2^2}$ and $\sqrt{x_3^2
- \hbar^2 \lambda_3^2}$.

Now let us consider active and passive transformations of
coordinates in 8-dimensional space of signals (\ref{s}). With a
passive transformation we mean a change of the coordinates $t,
x_n, \lambda_n$ and $\omega$, as opposed to an active
transformation which changes the basis $1,J_n,j_n$ and $I$.

The passive transformations of the octonionic coordinates $t, x_n,
\lambda_n$ and $\omega$, which leave invariant the norm
(\ref{sN}) form just $SO(4,4)$, obviously. We can represent these
transformations  of (\ref{s}) by the left products
\begin{equation}
s' = R s~,
\end{equation}
where $R$ is one of (\ref{R(j)}), (\ref{R(J,I)}). The operator $R$
simultaneously transform four planes of $s$. However, in three
planes $R$ can be rotate out by the proper choice of octonionic
basis. Thus $R$ can represent passive independent rotations in
four orthogonal planes of $s$ separately. Similarly we have some
four angles for other six operators (\ref{R(j)}), (\ref{R(J,I)})
and thus totally $4 \times 7 = 28$ parameters corresponding to
$SO(4,4)$ group of passive coordinate transformations. For
example, in the case of the decomposition (\ref{s-j1}) we can
introduce four arbitrary angles $\phi_t, \phi_x, \phi_\lambda$ and
$\phi_\omega$ and
\begin{equation} \label{s-j1-R}
s' = N_t e^{j_1 (\theta_t + \phi_t)} + N_x e^{j_1 (\theta_x+
\phi_x)}J_3  + N_\lambda e^{j_1(\theta_\lambda + \phi_\lambda)}j_2
+ N_\omega e^{j_1 (\theta_\omega + \phi_\omega)}I ~.
\end{equation}
Obviously under this transformations the norm (\ref{sN}) is
invariant. By the fine tuning of the angles in (\ref{s-j1-R}) we
can define rotations in any single plane from four.

Now let us consider active coordinate transformations, or
transformations of eight octonionic basis units $1, J_n, j_n$ and
$I$. For them, because of non-associativity, result of two
different rotations (\ref{R(j)}) and (\ref{R(J,I)}) are not
unique. This means that not all active octonionic transformation
generated by (\ref{R(j)}) and (\ref{R(J,I)}) form a group and can
be considered as a real rotation. Thus in the octonionic space
(\ref{s}) not to the all passive $SO(4,4)$-transformations we can
correspond active ones, only the transformations that have
realization with associative multiplications should be considered.
It was found that these associative transformations can be done by
the combine rotations of special form in at least two octonionic
planes. This kind of rotations also form a group (subgroup of
$SO(4,4)$), known as the automorphism group of split octonions
$G_2^{NC}$ (the real non-compact form of Cartan's exceptional Lee
group $G_2$). Some general results on $G_2^{NC}$ and its subgroup
structure one can find in \cite{BHW}.

Let us remind that the automorphism $A$ of a algebra is defined as
the transformations of the hyper-complex basis units $x$ and $y$
under which multiplication table of the algebra is invariant, i.e.
\begin{eqnarray} \label{A}
A(x+y) = Ax + Ay~, \nonumber \\
(Ax)(Ay) = (A(xy))~.
\end{eqnarray}
Associativity of this transformations is obvious from the second
relation and the set of all automorphisms of composition algebras
form a group. In the case of quaternions, because of
associativity, active and passive transformations, $SU(2)$ and
$SO(3)$ respectively, are isomorphic and quaternions are useful to
describe rotations in 3-dimensional space. There is different
situation for octonions. Each automorphism in the octonionic
algebra is completely defined by the images of three elements that
are not form quaternionic subalgebra, or all are not lay on the
same David's lime \cite{Zorn}. Consider one such set, say $(j_1,
j_2, J_1)$. Then there exists an automorphism such that
\begin{eqnarray} \label{auto}
j_1' &=& j_1~, \nonumber \\
j_2' &=& j_2\cos (\alpha_1 + \beta_1)/2 + j_3\sin (\alpha_1 + \beta_1)/2 ~, \\
J_1' &=& J_1\cos \beta_1 + I\sin \beta_1  ~, \nonumber
\end{eqnarray}
where $\alpha_1$ and $\beta_1$ are some independent real angles.
By definition (\ref{A}) automorphism does not affect unit scalar
$1$. The images of the other basis elements under automorphism
(\ref{auto}) are determined by the conditions:
\begin{eqnarray} \label{induced}
j_3' &=& j_1'j_2' = j_3\cos (\alpha_1 + \beta_1)/2 - j_2\sin (\alpha_1 + \beta_1)/2 ~, \nonumber \\
I'   &=& J_1'j_1'  = I \cos \beta_1 - J_1 \sin \beta_1~, \nonumber \\
J_2' &=& j_2'I' = J_2\cos (\alpha_1 - \beta_1)/2  + J_3\sin (\alpha_1 - \beta_1)/2 ~,  \\
J_3' &=& j_3'I' = J_3\cos (\alpha_1 - \beta_1)/2  - J_2\sin (\alpha_1 - \beta_1)/2 ~. \nonumber
\end{eqnarray}
It can easily be checked that transformed basis $J_n', j_n', I'$
satisfy the same multiplication rules as $J_n, j_n, I$.

There exists similar automorphisms with fixed $j_2$ and $j_3$
axis, which are generated by the angles $\alpha_2, \beta_2$ and
$\alpha_3, \beta_3$ respectively.

One can define also hyperbolic automorphisms for the vector-like
units $J_n$ by the angles $u_n,k_n$. For example, if we fix the
axis $J_1$ then corresponding to (\ref{auto}) and (\ref{induced})
transformations are
\begin{eqnarray}
J_1' &=& J_1~, \nonumber \\
J_2' &=& J_2 \cosh (k_1 + u_1)/2 + j_3 \sinh (k_1 + u_1)/2 ~, \nonumber\\
I'   &=& I \cosh u_1 - j_1 \sinh u_1  ~, \nonumber \\
j_3' &=& J_1'J_2' = j_3\cosh (k_1 + u_1)/2 + j_2\sinh (k_1 + u_1)/2 ~, \\
j_1' &=& J_1'I'  = j_1 \cosh u_1 - I \sinh u_1~, \nonumber \\
j_2' &=& J_2'I' = j_2\cosh (k_1 - u_1)/2  + J_3 \sinh (k_1 - u_1)/2 ~, \nonumber \\
J_3' &=& j_3'I' = J_3\cosh (k_1 - u_1)/2 + J_2\sinh (k_1 - u_1)/2 ~. \nonumber
\end{eqnarray}

Similarly automorphism with the fixed seventh axis $I$ has the form:
\begin{eqnarray}
I'   &=& I~, \nonumber \\
j_1' &=& j_1 \cosh \sigma_1 + J_1 \sinh \sigma_1 ~, \nonumber\\
j_2' &=& j_2 \cosh \sigma_2 + J_2 \sinh \sigma_2  ~, \nonumber \\
J_1' &=& j_1'I' = J_1\cosh \sigma_1 + j_1\sinh \sigma_1 ~, \\
J_2' &=& j_2'I'  = J_2 \cosh \sigma_2 + j_2 \sinh \sigma_2 ~, \nonumber \\
j_3' &=& j_1'j_2' = j_3 \cosh (\sigma_1 + \sigma_2) - J_3 \sinh (\sigma_1 + \sigma_2) ~, \nonumber \\
J_3' &=& j_3'I' = J_3 \cosh (\sigma_1 + \sigma_2) - j_3\sinh (\sigma_1 + \sigma_2) ~. \nonumber
\end{eqnarray}

So for each octonionic basis there are seven independent
automorphism each introducing two angles that correspond to $2
\times 7 = 14$ generators of the algebra $G_2^{NC}$. For our
choice of octonionic basis infinitesimal passive transformation of
the coordinates, corresponding to $G_2^{NC}$, can be written as
\begin{eqnarray} \label{x,l,o}
t' &=& t~, \nonumber \\
x_i' &=& x_i - \frac 12 \epsilon_{ijk} \left(\alpha^j -
\beta^j\right) x^k + c\hbar \beta^i \omega +
\frac \hbar 2 \left( U_{ik} - \epsilon_{ijk} u^j\right) \lambda^k~, \nonumber \\
\omega' &=& \omega - \frac 1{c\hbar}\beta_ix^i - \frac 1c u_i\lambda^i~, \\
\lambda_i' &=& \lambda_i -\frac 12 \epsilon_{ijk} \left(\alpha^j +
\beta^j\right) \lambda^k - c u^i \omega + \frac 1{2\hbar} \left(
U_{ik} + \epsilon_{ijk} u^j\right) x^k~, \nonumber
\end{eqnarray}
where $U_{ik}$ is the symmetric matrix
\begin{equation}
U =\pmatrix{2\sigma_1 & k_3 & k_2 \cr k_3 & 2\sigma_2 & k_1 \cr
k_2 & k_1 & -2(\sigma_1 + \sigma_2) }~.
\end{equation}

In the limit $(\hbar \lambda, \hbar \omega \rightarrow 0)$ the
transformations (\ref{x,l,o}) reduce to the standard $O(3)$
rotations of Euclidean 3-space by the Euler angles $\phi_n =
\alpha_n - \beta_n$. However, for some problems in quantum regime
extra symmetries can be retrieved.

The formulas (\ref{x,l,o}) represent rotations of (3,4)-sphere
that is orthogonal to the time coordinate $t$. To define the busts
note that active and passive form of mutual transformations of $t$
with $x_n, \lambda_n$ and $\omega$ are isomorphic and can be
described by the seven operators (\ref{R(j)}) and (\ref{R(J,I)})
(e.g. first term in (\ref{s-j1-R})), which form the group
$O(3,4)$. In the case $(\hbar \lambda, \hbar \omega \rightarrow
0)$ we stay with the standard $O(3)$ Lorentz boost in the
Minkowski space-time governing by the operators $e^{J_n m_n }$,
where $m_n = \arctan v_n/c$.


\section{Conclusion}

In this paper the David's Star shape duality plane that describe
multiplications table of basis units of split octonions (instead
of the Fano triangle of ordinary octonions) was introduced.
Different kind of rotations in the split octonionic space were
considered. It was shown that in octonionic space active and
passive transformations of coordinates are not equivalent. The
group of passive coordinate transformations, which leave invariant
the norms of split octonions, is $SO(4,4)$, while active rotations
are done by the direct product of the seven $O(3,4)$-boosts and
fourteen $G_2^{NC}$-rotations. In classical limit these
transformations give the standard 6-parametrical Lorentz group.


\end{document}